\documentclass[a4paper,12pt]{article}
\usepackage{jheppub}

\usepackage[applemac]{inputenc}
\usepackage{amsmath}
\usepackage{slashed}
\usepackage{amscd,amstext,amsbsy,amsopn,amsthm,amsxtra,upref,amsfonts,amssymb,euscript,latexsym,graphicx,color}
\usepackage[all]{xy}
\usepackage[american]{babel}
\usepackage{multicol}
\usepackage{tikz}
\usepackage{tikz-3dplot}
\usepackage{mdframed}
\usepackage{todonotes}
\usetikzlibrary{patterns}
\usepackage{hyperref,pgfplots,mdframed,subcaption}
\usepackage[T1]{fontenc}

\usepackage{mathrsfs}

\usepackage[all]{xy}
\usepackage{tikz-cd}
\usepackage{graphicx}

\usepackage{amsmath,bbm,array,cleveref,amsfonts,subcaption,graphicx,wrapfig,lscape,youngtab,bbold,float,multirow,longtable,rotating,epstopdf}

\tikzstyle{white}=[circle,draw,
inner sep=0pt,minimum size=2mm]
\tikzstyle{black}=[circle,fill=black,inner sep=0pt,minimum size=2mm]

\def\beqa{\begin{eqnarray}}
\def\eeqa{\end{eqnarray}}

\newcommand{\drawsquare}[2]{\hbox{%
\rule{#2pt}{#1pt}\hskip-#2pt
\rule{#1pt}{#2pt}\hskip-#1pt
\rule[#1pt]{#1pt}{#2pt}}\rule[#1pt]{#2pt}{#2pt}\hskip-#2pt
\rule{#2pt}{#1pt}}

\newcommand{\fund}{~\raisebox{-.5pt}{\drawsquare{6.5}{0.4}}~}
\newcommand{\antifund}{~\overline{\raisebox{-.5pt}{\drawsquare{6.5}{0.4}}}~}

\newcommand{\asymm}{~\raisebox{-3.5pt}{\drawsquare{6.5}{0.4}}\hskip-6.9pt%
        \raisebox{3pt}{\drawsquare{6.5}{0.4}}~}

\newcommand{\antiasymm}{~\overline{\raisebox{-3.5pt}{\drawsquare{6.5}{0.4}}\hskip-6.9pt%
        \raisebox{3pt}{\drawsquare{6.5}{0.4}}}~}

\usepackage{booktabs}
\makeatletter
\newcommand*\variableheghtrulefill[1][.4\p@]{%
	\leavevmode
	\leaders \hrule \@height #1\relax \hfill
	\null
}
\makeatother

\title{
The Octagon at large M
}

\author[a]{Riccardo Argurio}
\author[b]{Matteo Bertolini}
\author[c,d,e]{Sebasti\'an Franco}
\author[a]{Eduardo Garc\'{\i}a-Valdecasas}
\author[f]{Shani Meynet}
\author[g]{Antoine Pasternak}
\author[h]{Valdo Tatitscheff}

\affiliation[a]{Physique Th\'eorique et Math\'ematique and International Solvay Institutes \\ Universit\'e Libre de Bruxelles; C.P. 231, 1050 Brussels, Belgium}

\affiliation[b]{SISSA and INFN, Via Bonomea 265; I 34136 Trieste, Italy}

\affiliation[c]{Physics Department, The City College of the CUNY \\ 160 Convent Avenue, New York, NY 10031, USA}

\affiliation[d]{Physics Program and $^e$Initiative for the Theoretical Sciences \\
The Graduate School and University Center, The City University of New York  \\
365 Fifth Avenue, New York NY 10016, USA}

\affiliation[f]{Mathematics Institute, Uppsala University, Box 480, SE-75106 Uppsala, Sweden}

\affiliation[g]{Stanford Institute for Theoretical Physics, Stanford University, Stanford, CA, 94305}

\affiliation[h]{IRMA, UMR 7501, Universit\'e de Strasbourg et CNRS \\ 
7 rue Ren\'e Descartes 67000 Strasbourg, France}

\emailAdd{rargurio@ulb.ac.be, bertmat@sissa.it, sfranco@ccny.cuny.edu, eduardo.garcia.valdecasas@gmail.com, shanimeynet@hotmail.it, antoinepasternak14@gmail.com, valdotatitscheff@gmail.com}

\abstract{Recently, the first instance of a model of D-branes at Calabi-Yau singularities where supersymmetry is broken dynamically into stable vacua has been proposed. This construction was based on a system of  $N$ regular and $M=1$ fractional branes placed at the tip of the so-called (orientifolded) Octagon singularity. In this paper we show that this model admits a large $M$ generalization, having the same low energy effective dynamics. This opens up the possibility that the effect on geometry is smooth, and amenable to describing the gauge theory all along the RG flow, including the deep IR, in terms of a weakly coupled gravity dual background. The relevance of this result in the wider context of the string landscape and the Swampland program is also discussed.}

\begin{document}
	\maketitle

\section{Introduction}
\label{intro}

Since the early days of the AdS/CFT correspondence many efforts were made to extend it beyond the realm of conformal field theories. Set-ups consisting of D-branes at Calabi-Yau (CY) singularities have been extremely successful in this program. When the D-brane bound state includes both regular and fractional branes the dual gauge theory is an ${\cal N}=1$ or ${\cal N}=2$ supersymmetric theory enjoying a non-trivial RG flow. This way, several field theory phenomena have found a dual supergravity (or string) description, providing us with a novel perspective on the strong coupling regime of gauge theories. It is an obvious question to ask whether {\it any} possible low-energy behavior can be found using this approach.

While confinement, dynamical mass generation, chiral symmetry breaking as well as other phases that ${\cal N}=1$ and ${\cal N}=2$ gauge theories can exhibit at low energy have found a dual description, theories breaking supersymmetry dynamically did not find any reliable one until recently. The only known examples were models in which supersymmetry was broken in metastable vacua \cite{Kachru:2002gs,Argurio:2006ny,Argurio:2007qk} or in a runaway fashion \cite{Berenstein:2005xa,Franco:2005zu,Bertolini:2005di,Intriligator:2005aw}. 

Recently, this gap has been filled and the first example of a CY singularity which can host D-brane bound states triggering dynamical supersymmetric breaking (DSB) into stable vacua was found, the so-called Octagon model \cite{Argurio:2020dkg,Argurio:2020npm}. The D-brane configuration consists of $N$ regular and one fractional branes on top of an orientifold plane placed at the tip of a CY cone. The gauge theory enjoys a duality cascade, very much like the Klebanov-Strassler model \cite{Klebanov:2000hb} and many of its siblings. Along the RG flow, the effective number of regular branes decreases and in the deep IR the theory reduces to a double copy of the well-known $SU(5)$ model \cite{Affleck:1983vc}, which breaks supersymmetry dynamically in a stable vacuum. A careful analysis of the moduli space excludes the existence of lower energy vacua somewhere else in the space of field VEVs and this guarantees the stability of the DSB vacuum. This shows that DSB into stable vacua can be realized in D-brane models and hence within the gauge/string duality. 

One still unsatisfactory aspect of the analysis of \cite{Argurio:2020dkg,Argurio:2020npm} is that the gauge theory at hand would not admit a weakly coupled gravity dual description all along the RG flow, since in the deep IR, where the number of regular branes becomes small, the curvature would be large in string units, since one is left with order one fractional branes.  

In this work we show that the Octagon model admits a large $M$ generalization, obtained by replacing the only fractional brane with $M$ such branes. The gauge theory at the bottom of the cascade is now more complicated. It is a gauged version of a double copy of an $SU(N)$ supersymmetric gauge theory with one antisymmetric tensor and $N-4$ flavors in the antifundamental, which was shown in \cite{Affleck:1984xz} to break supersymmetry dynamically for odd $N$. In fact, in our case the IR dynamics of the brane system is described by a model with three gauge group factors, $G= SU(N)_1 \times SU(N-4)_2 \times SU(N)_3$, since the flavor group, shared by the two $SU(N)$ factors, is now gauged. Via a detailed analysis of the classical and quantum moduli space, we show that the theory at the bottom of the cascade still breaks supersymmetry in a stable vacuum, in a similar fashion as the original model of \cite{Affleck:1984xz}. As we will see, $N$ is related to $M$ as $N=M+4$, and stabilization occurs only if the number of fractional branes is odd. The stability of the supersymmetry breaking vacuum  crucially depends on the presence of a tree-level sextic superpotential which couples chiral fields charged under all gauge groups and lifts all classical flat directions. What is remarkable is that all ingredients to make the otherwise runaway vacuum stable (matter content, gauge group ranks and, most notably, the sextic superpotential) come automatically out of the string construction, without any external input.  

This shows that the Octagon singularity can host a D-brane model triggering DSB into stable vacua which can allow for a weakly coupled dual description all along the flow. After the geometric transition, the curvature is proportional to $1/(\alpha' g_s M)$ in the IR, and hence small in string units for $M \gg 1$.  

The rest of the paper is organized as follows. 
In section \ref{review Oct} we review the Octagon model and its stability for $M=1$, following  \cite{Argurio:2020dkg}.  In section \ref{DSB_M}, which is the main part of the paper and is purely field theoretical, we show that supersymmetry is dynamically broken and the vacuum stable for any odd $M$. We also show that for even $M$ there is always a runaway. In section \ref{geomdual}, we discuss the geometric dual side of the story and compare the Octagon to previous models where supersymmetry breaking was found to be runaway, {\it e.g} \cite{Berenstein:2005xa,Franco:2005zu,Bertolini:2005di}. In particular, we discuss what the geometric mechanism stabilizing the otherwise runaway vacuum possibly is. Finally, in section \ref{disc}, we discuss to what extent our results might be relevant in the wider context of the Swampland program and in de Sitter constructions in string theory.

\section{Review of the Octagon model}
\label{review Oct}

In this section we review the Octagon model. We refer the reader to \cite{Argurio:2020dkg,Argurio:2020npm} for more details.

In the context of models of D-branes at CY singularities, evidence has been given in \cite{Franco:2007ii} that the only instance where models of dynamical supersymmetry breaking could possibly be accommodated were orientifold models. They can be engineered in terms of fractional branes and in this way the simplest DSB models, namely the $SU(5)$ and the 3-2 models, were realized \cite{Buratti:2018onj,Argurio:2019eqb}. However, it turned out that in the decoupling limit, where the fractional brane system is part of a UV-complete large $N$ model, such brane configurations become actually unstable and supersymmetric vacua exist somewhere else in field space. In particular, in all models an instability associated to a local non-isolated singularity at the tip of the CY cone exists, and the brane system relaxes to a supersymmetry preserving vacuum \cite{Argurio:2019eqb}. In a subsequent work \cite{Argurio:2020npm}, exploiting dimer techniques, a thorough analysis was pursued on geometric properties of toric CYs with orientifolds and it was possible to elucidate the basic properties a CY should have in order to host a DSB model and be free of non-isolated singularities, hence evading the aforementioned no-go theorem. This singled-out the Octagon as the simplest such CY.

\begin{figure}[h!]
				\centering
				\includegraphics[scale=0.25]{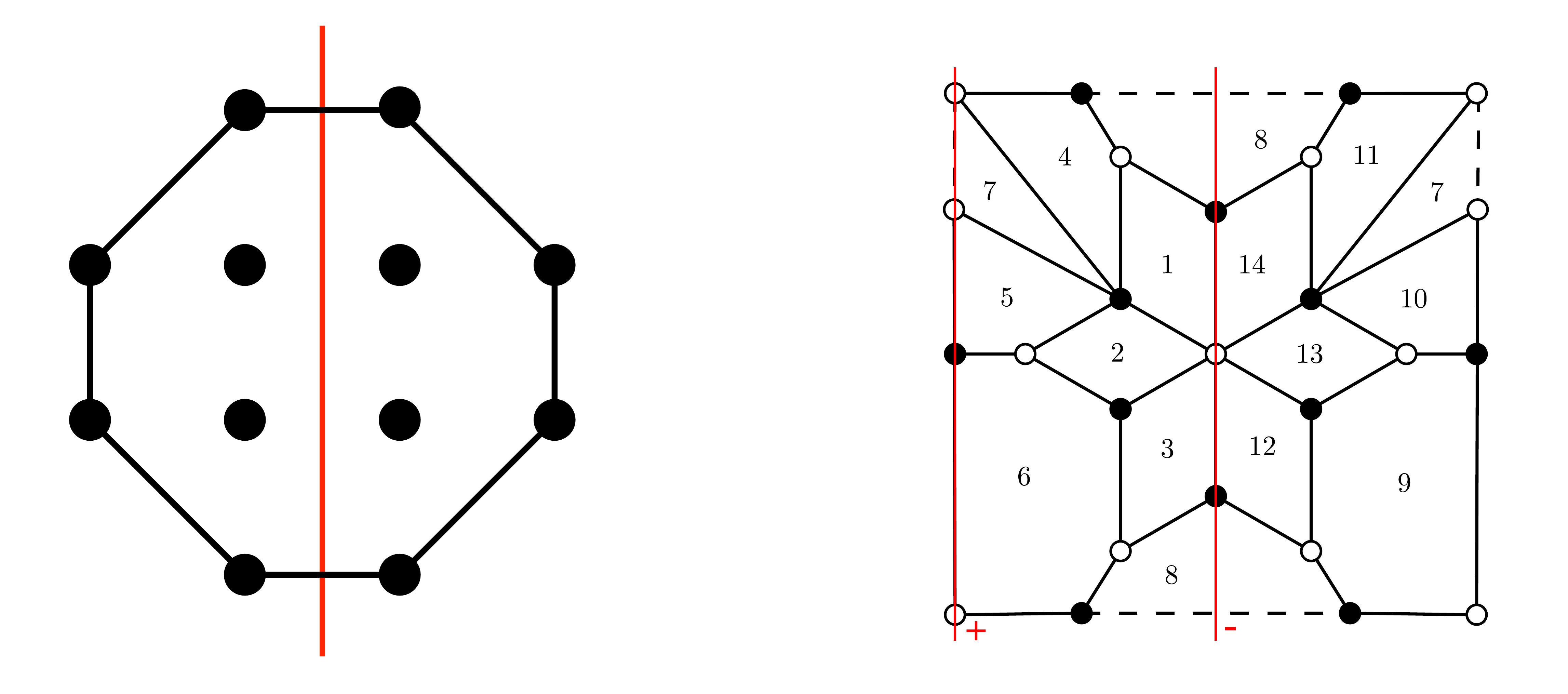}
				\caption{On the left the toric diagram of the Octagon singularity. On the right the dimer representing one of its toric phases. Red lines represent the orientifold planes and the $\pm$ signs their charges. The orientifold projection identifies faces (1,..., 6) with faces (14,..., 9) while faces 7 and 8 get self-identified and become $SO$ and $USp$ factors, respectively.}
				\label{Fig:Octagon T+D}
	\end{figure}
	
In figure \ref{Fig:Octagon T+D} we present the toric diagram of the Octagon and the dimer corresponding to one of its toric phases, one which makes the orientifold action manifest. The Octagon has a rich singularity structure and admits five independent types of deformation fractional branes.\footnote{We adhere to the terminology introduced in \cite{Franco:2005zu}, where different classes of fractional branes were defined according to the low energy dynamics they trigger: deformation (leading to confinement), ${\cal N}=2$ and DSB fractional branes.} The one relevant to us populates faces 1, 2, 3 and 7 of the dimer (and their mirrors under the orientifold involution).

The orientifold gauge theory, obtained placing $N$ regular and $M$ such fractional branes at the conical singularity, has gauge group
\beqa
G &=& SU(N+M+4)_1 \times SU(N+M)_2 \times SU(N+M+4)_3 \times SU(N)_4 \times \nonumber \\ 
&& \times SU(N)_5 \times SU(N)_6 \times SO(N+M+4)_7 \times USp(N)_8 ~,
\label{Ggroup}
\eeqa
matter in bifundamental, symmetric and antisymmetric representations and a superpotential with up to sextic terms. Note that in quiver gauge theories (and the Octagon is no exception) orientifolds typically contribute undemocratically to the gauge group ranks, very much like fractional branes do. Relevant to what we will discuss in section \ref{geomdual} is to note, see \eqref{Ggroup}, that the orientifold we consider contributes to the ranks of groups at faces 1, 3 and 7, while the deformation fractional branes to groups at faces 1, 2, 3, and 7. This indicates that the fractional brane and the orientifold plane are not fully aligned, geometrically.  

The theory \eqref{Ggroup} undergoes a complicated RG flow, which can be interpreted in terms of a duality cascade. As shown in \cite{Argurio:2020dkg,Argurio:2020npm}, in the IR the effective dynamics reduces to that of a three gauge groups model, $G= SU(M+4)_1 \times SU(M)_2 \times SU(M+4)_3$, with charged matter and a single sextic superpotential term (plus a decoupled, $SO(M+4)$ pure SYM node which confines on its own and plays no role in the rest of the discussion). For $M=1$ the low energy effective dynamics is the same as a double copy of the $SU(5)$ DSB model of \cite{Affleck:1983vc}, dubbed Twin $SU(5)$ model in \cite{Argurio:2020dkg}. This case is simple to analyze since the middle gauge factor, $SU(M)_2$, becomes trivial, the superpotential vanishes and one is left with two copies of the $SU(5)$ model, which individually break supersymmetry. The analysis for $M>1$ is more involved, but potentially far-reaching, as emphasized in section \ref{intro}. This is what we will focus on in section \ref{DSB_M}. 

An analysis of all (known) possible decay channels was performed in \cite{Argurio:2020dkg}, showing that no instabilities affect the model. This excludes the existence of supersymmetric vacua elsewhere in the space of field VEVs or runaway to infinity. In particular, the Octagon is an isolated singularity and as such it avoids the no-go theorem of \cite{Argurio:2019eqb}. Indeed, non-isolated singularities have toric diagrams with points inside the edges along the boundary, which the Octagon does not display, see figure \ref{Fig:Octagon T+D}. 

\begin{figure}[h!]
				\centering
				\includegraphics[scale=0.30]{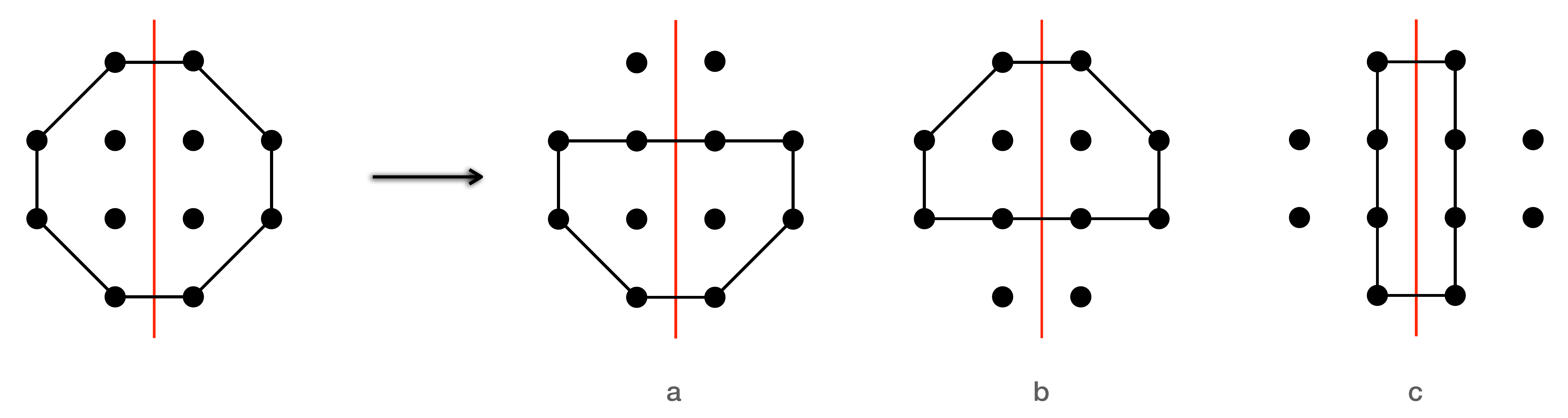}
				\caption{Examples of partial resolutions of the Octagon singularity. In all cases, a local non-isolated singularity shows up in the partially resolved singularity.}
				\label{Fig:Octagon PR}
	\end{figure}

Another source of potential instability comes from partial resolutions of the singularity which correspond to non-vanishing baryonic VEVs in the dual field theory \cite{Morrison:1998cs,Feng:2000mi,Garcia-Etxebarria:2006ngz}. The partially resolved singularity could have (local) non-isolated singularities, making the baryonic branch unstable, following again the no-go theorem of  \cite{Argurio:2019eqb}. The Octagon is no exception and some of such partial resolutions are shown in figure \ref{Fig:Octagon PR}. In the presence of fractional branes, however, partial resolutions may be obstructed \cite{Garcia-Etxebarria:2006ngz}. A simple way to see how this happens is to note that from the dimer viewpoint partial resolutions correspond to the fusion of adjacent faces in the dimer. For this to happen, the ranks of the gauge groups associated to such faces must be equal. It turns out that this cannot happen in our model, due to rank unbalance of the corresponding adjacent faces, either because $M \not =0$ or due to the orientifold. This implies that  all dangerous partial resolutions are obstructed and this kills any possible ${\cal N}=2$-like instabilities along the baryonic branch. 

\section{Dynamical supersymmetry breaking at large M}
\label{DSB_M}

The model that we want to analyze describes the IR dynamics of a system of $N$ regular and $M$ fractional branes at the (orientifolded) Octagon singularity \cite{Argurio:2020dkg}. The gauge group is $SU(M+4)_1 \times SU(M)_2 \times SU(M+4)_3$ with matter content
	\begin{equation}
	A_{11} = \asymm_1 \, , \quad X_{12} = (\antifund_1, \fund_2) \, , \quad  X_{23} = (\antifund_2, \fund_3) \, , \quad A_{33} = \antiasymm_3 \, , \label{Eq:FieldsRep}
	\end{equation}
	and superpotential
	\begin{equation}
	W = h \,  \mathrm{tr}A_{11}X_{12}X_{23}A_{33} X_{23}^T X_{12}^T \, , \label{Eq:W}
	\end{equation}
	where $h$ is an holomorphic coupling. 
		\begin{figure}[h!]
				\centering
				\includegraphics[scale=0.45]{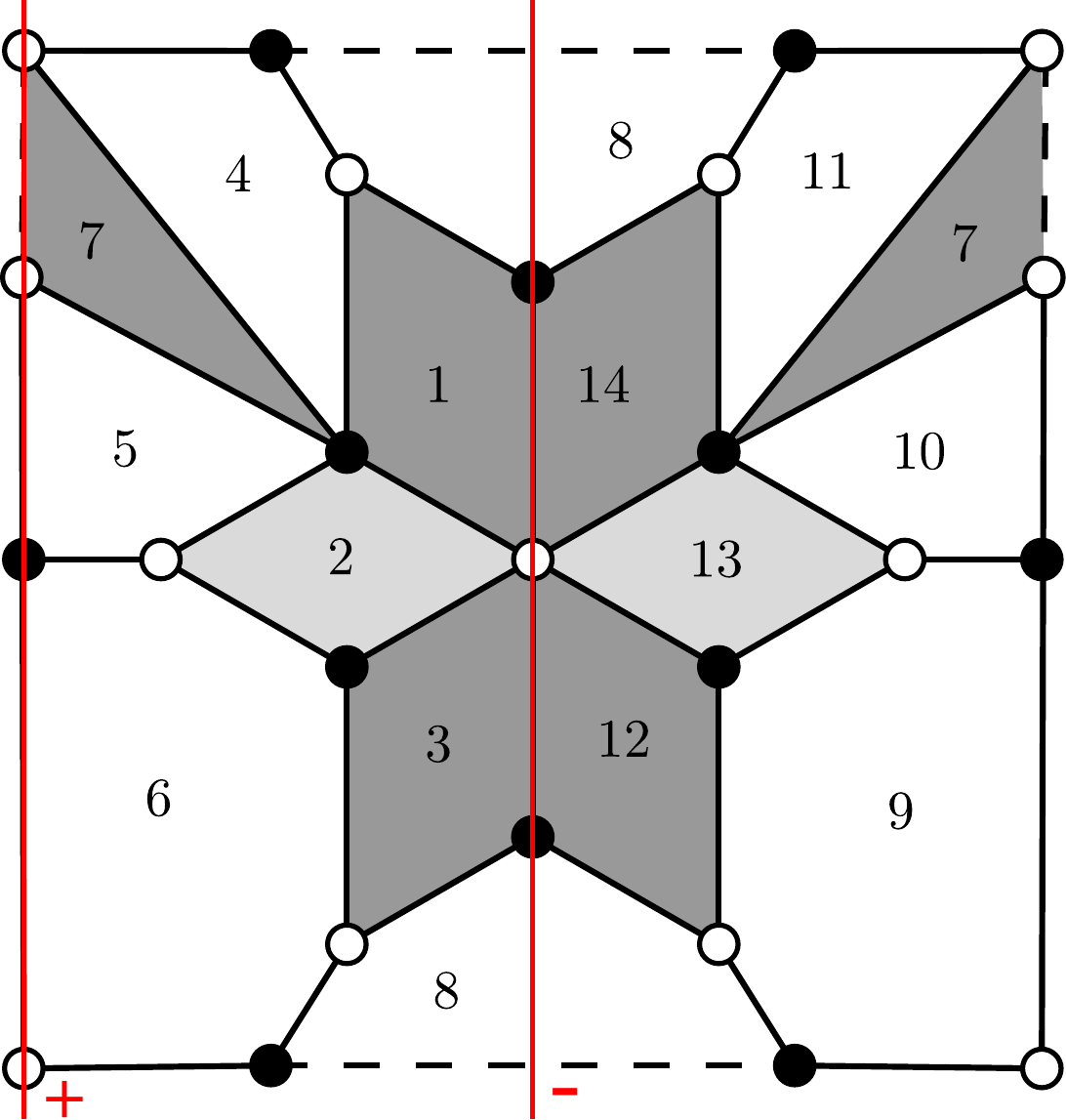}
				\caption{The Octagon model with $N$ regular and $M$ fractional branes and orientifold planes (represented by the red lines). White faces correspond to gauge factors to which only the $N$ regular branes contribute, grey faces get instead contribution also from the $M$ fractional branes. Dark grey correspond to faces where also orientifold charge contributes, see eq.~\eqref{Ggroup}.}
				\label{Fig:Octagon}
	\end{figure}

	If $SU(M)_2$ were not gauged this model would just be a double copy  of the model considered in \cite{Affleck:1984xz}, which was shown to break supersymmetry dynamically when $M$ is odd and is runaway when $M$ is even. Using the same techniques, we will show that the same happens when $SU(M)_2$ is gauged, even though the gauge theory is way richer and the dynamics more complicated. 	
	
We will start by analyzing the theory at the classical level and then consider non-perturbative effects. We will assume $M$ to be odd below. In section \ref{section_even_M} we will show what changes in the even $M$ case.

\subsection{Classical supersymmetric vacuum}
	
The moduli space is parametrized in terms of chiral gauge invariants, constructed from the UV fields
	\begin{equation}
		(A_{11})^{ab}  \, , \quad (X_{12})_a^{\,\,\, i}  \, , \quad  (X_{23})_i^{\, m}  \, , \quad (A_{33})_{mn}  ~ , 
		\label{Eq:FieldsInd}
	\end{equation}
where we distinguish the color indices for the different factors of the gauge group as follows: $a,b,c$ for $SU(M+4)_1$, $i,j,k$ for  $SU(M)_2$, and $m,n,p$ for $SU(M+4)_3$. 
		
In order to construct all gauge invariant operators, it is convenient to begin by contracting the indices of $SU(M)_2$. This step is equivalent to treating this gauge group as $SU(M)$ SQCD with $N_f = M+4$ flavors and building its corresponding gauge invariants in the standard way. We therefore have mesonic and baryonic operators, given by
	\begin{equation}
		\begin{array}{rcl}
			(\mathcal{M})_a^{\,\,\, m} &=& (X_{12})_a^{\,i} (X_{23})_i^{\,m} \, , \\[.1 cm]
			(\mathcal{B}_{1})_{\lbrack a_1 \cdots a_M\rbrack} &=& \epsilon_{i_1 \cdots i_{M}}(X_{12})_{a_1}^{\,i_1}\cdots (X_{12})_{a_M}^{\,i_M} \, , \\[.1 cm]
			(\mathcal{B}_{3})^{\lbrack m_1 \cdots m_M\rbrack} &=& \epsilon^{i_1 \cdots i_{M}}(X_{23})_{i_1}^{\, m_1}\cdots (X_{23})_{i_M}^{\, m_M} \, .
		\end{array}
		\label{mesons_baryons_SU(M)_2}
	\end{equation}
The $\lbrack\cdots \rbrack$ notation for baryonic operators indicates the antisymmetrization of the indices in brackets. Note that using $\epsilon_{i_1 \cdots i_{M}}\epsilon^{j_1 \cdots j_{M}}=M!\  \delta^{[j_1}_{i_1}\dots \delta^{j_M]}_{i_M}$, one can show that $\mathcal{B}_{1}\mathcal{B}_{3}$ is expressed as a sum of products of $\mathcal{M}$ matrices. 

Let us now consider how to use the operators in \eqref{mesons_baryons_SU(M)_2} as building blocks for invariants under the remaining gauge groups. Focusing on $SU(M+4)_1$ first, we note that
	\begin{equation}
		(\mathcal{B}_{1})_{\lbrack a_1 \cdots a_M\rbrack} \epsilon^{a_1 \cdots a_M a_{M+1} \cdots a_{M+4}} (\mathcal{M})_{a_{M+1}}^{\,\,\, m} = 0 \, .
	\end{equation}
Then, since $\mathcal{B}_1$ and $A_{11}$ have an odd and even number of indices under this gauge group, respectively, it is therefore impossible to construct gauge invariant operators that simultaneously include $\mathcal{B}_1$ and $A_{11}$. This, in turn, implies that $\mathcal{B}_1$ is not helpful for building invariants for all gauge groups. An identical argument applies to $SU(M+4)_3$ and implies that, similarly, $\mathcal{B}_3$ cannot be used to construct gauge invariants. From this analysis, we conclude that gauge invariants must be the trace of (powers of)
	\begin{equation}
	(\mathcal{X})^{a}_{\, \, \, b} = (A_{11})^{ac}(\mathcal{M})_{c}^{\, m} (A_{33})_{mn} (\mathcal{M}^T)_{\, b}^n ~ ,
	\end{equation}
Indeed, the superpotential \eqref{Eq:W} is given by
\begin{equation}
		W = h \, \mathrm{tr} \mathcal{X} \, .
		\label{W_1}
	\end{equation}

Let us now consider the gauge invariants associated to higher powers of $\mathcal{X}$,
	\begin{equation}
		\mathrm{tr} \mathcal{X}^k \, , \label{Eq:GaugInv}
	\end{equation}
	for an integer $1\leqslant k\leqslant (M-1)/2$. The upper bound on $k$, {\it i.e.} the finite number of operators, follows from the fact that $\mathcal{X}$ can be written as a product of $\mathcal{A} = X_{12}^T A_{11} X_{12}$ and $\tilde{\mathcal{A}} = X_{23} A_{33} X_{23}^T$, two antisymmetric matrices of maximal rank $M-1$ for $M$ odd. Moreover, antisymmetric matrices have degenerate eigenvalues of multiplicities two, which leads to $(M-1)/2$ distinct eigenvalues accessible with the operators in \eqref{Eq:GaugInv}.

\paragraph{Effect of non-zero superpotential.}
The $(M-1)/2$ operators \eqref{Eq:GaugInv} parametrize the space of the D-flat directions which, in the absence of the superpotential, is nothing but the moduli space. However, the superpotential \eqref{Eq:W} gives rise to the following $F$-flatness conditions
	\begin{equation}
	\begin{array}{rlcrl}
	(X_{12}X_{23}A_{33}X_{23}^TX_{12}^T)_{ab} & = 0 \, , & \quad \quad & (X_{23}A_{33}X_{23}^TX_{12}^T A_{11})_{i}^{\, \, \, a} & = 0 \, , \\[.1cm]
	(A_{33}X_{23}^TX_{12}^T A_{11} X_{12})_{m}^{\,\,\, i} & = 0 \, , & \quad \quad &	 (X_{23}^TX_{12}^T A_{11}X_{12}X_{23})^{mn} & = 0 \, .
	\end{array}
	\end{equation}
They imply the vacuum equation
	\begin{equation}
		(\mathcal{X})^a_{\,\,\, b}  = 0 \, ,
	\end{equation}
which in turn leads to
	\begin{equation}
		\left< \mathrm{tr} \mathcal{X}^k \right> = 0  \, , 
	\end{equation}
	for every $k$. The upshot is that  the superpotential lifts all D-flat directions and the classical moduli space is reduced to a single point, the origin.
	
\subsubsection*{Dynamics along the D-Flat directions} 
\label{Sec:DFlat}

In this section we would like to repeat the above analysis in terms of  elementary UV fields. This is somehow redundant and, not surprisingly, the results will be fully consistent with the computations using gauge invariant variables. What we are actually interested in is the symmetry breaking pattern and IR behavior of the theory along the D-flat directions, something that an analysis in term of gauge invariant operators obscures. Having such an understanding is important for what will come next, namely the study of quantum corrections which, as we will discuss in section \ref{np_eff1}, drive the vacuum away from the origin of field space. Below we summarize the main conclusions and refer to Appendix \ref{section_low_E_theory_generic_vevs} for details. 

Let us start assuming that the superpotential vanishes and investigate the theory along the D-flat directions. The D-flatness conditions read
	\begin{align}
SU(M+4)_1  :  \ \ \ & 2(A_{11})^{bc}(A_{11}^\dagger)_{ca} -  (X_{12}^\dagger)^{\, b}_{i}(X_{12})^{\, i}_{ a} =   \alpha_1  \, \delta^b_{\, a} \, , \label{Eq:Dterm1} \\
SU(M)_2	 :  \ \ \ &  (X_{12})^{\,\, j}_{ a}(X_{12}^\dagger)^{\, a}_{i} - (X_{23}^\dagger)^{\,\, j}_{m}(X_{23})^{\, m}_{ i}  =  \alpha_2  \, \delta^j_{\, i} \, , \label{Eq:Dterm2} \\
SU(M+4)_3 : \ \ \ & (X_{23})^{\, n}_{i}(X_{23}^\dagger)^{\,\, i}_{ m} - 2(A_{33}^\dagger)^{np}(A_{33})_{pm}  =  \alpha_3  \, \delta^n_{\, m} \, . \label{Eq:Dterm3}
	\end{align}
Using  (the global part of the) gauge and global symmetries the solutions to these equations can be simplified. The upshot is that the D-flat directions can be parametri\-zed in terms of $(M-1)/2$ parameters $v_i$ (which, consistently, equal the number of gauge invariant operators $\mathrm{tr} \mathcal{X}^k$ in \eqref{Eq:GaugInv}). 

One can then analyze the low energy dynamics along such D-flat directions as a function of the VEVs. When all the $v_i$ are non-zero and different, the gauge theory reduces to the Twin $SU(5)$ model plus $(M-1)/2$ decoupled $SU(2)\simeq USp(2)$ pure SYM sectors, and $(M-1)/2$ singlets 
\begin{equation}
\begin{array}{ccccccc}
	\begin{array}{c}
	SU(5)_1  \\
	\asymm_1 \, , \quad \antifund_1 
	\end{array} & \boldsymbol{\times} & \begin{array}{c} SU(5)_3 \\
  \fund_3\, , \quad \antiasymm_3
\end{array} & \boldsymbol{\times} &  \begin{array}{c}
	USp(2)^{\frac{M-1}{2}} \\
	\text{pure SYM} 
\end{array} & \boldsymbol{+} & \frac{M-1}{2} \text{ singlets.}
\end{array}
\label{theory_pf_directions_1}
\end{equation}
When all $v_i$ are equal, the $USp(2)$ factors combine into an enhanced $USp(M-1)$ gauge group and all but one of the singlets enhance into an antisymmetric representation. One then obtains 
\begin{equation}
	\begin{array}{ccccccc}
		\begin{array}{c}
		SU(5)_1  \\
		\asymm_1 \, , \quad \antifund_1 
	\end{array} & \boldsymbol{\times} & \begin{array}{c} SU(5)_3 \\
		\fund_3\, , \quad \antiasymm_3
	\end{array} & \boldsymbol{\times} &  \begin{array}{c}
			USp(M-1)_\text{diag} \\
		 \asymm
	\end{array} & \boldsymbol{+} &  \begin{array}{c}
	\text{one singlet.}
	\end{array} 
	\end{array}
	\label{theory_pf_directions_2}
\end{equation}
It is straightforward to see that when subsets of the $v_i$ are equal, one gets intermediate theories between \eqref{theory_pf_directions_1} and \eqref{theory_pf_directions_2}.

Let us now consider the effect of adding the superpotential. In the case of all $v_i=v$, working to first order in the fluctuations around a given vacuum (see again Appendix \ref{section_low_E_theory_generic_vevs} for details and notation) the superpotential becomes
{\begin{equation}
W = h \, \mathrm{tr} \mathcal{X} = \frac{M-1}{2}h v^6 +\frac{15}{2}h v^4 \mathrm{tr}\mathbb{J}\delta A_\text{d}\mathbb{J}\delta A_\text{d} + \mathcal{O}(\delta A_\text{d}^3)\, ,
\label{perturbative_superpotential_pseudoflat_directions}
\end{equation}
where $v$ is the dynamical singlet VEV and $\delta A_\text{d}$ is the $\mathbb{J}$-traceless antisymmetric tensor of $USp$. Computing the scalar potential from this superpotential, one can see that it is minimized for $v=0$, for which the vacuum is indeed supersymmetric. One can also see that for any non-zero value of $v$, the field $\delta A_\text{d}$ is massive. The dynamics of $USp$ with a massive traceless antisymmetric tensor is the one of a pure SYM at low energies, hence it is supersymmetric and it does not contribute to the vacuum energy. A similar computation can be performed when not all $v_i$ are equal, for which the classical moduli is also found to be just the origin. Once again, this conclusion coincides with the analysis in the previous section.

\subsection{Non-perturbative effects and supersymmetry breaking} 
\label{np_eff1}

In this section we will analyze the fate of the classical supersymmetric vacuum located at the origin of field space once quantum corrections are taken into account. More precisely, we will study the low energy behavior of the model and determine properties of the vacuum when $M$ is odd and strictly greater than one. We will comment on the case of even $M$ in section \ref{section_even_M}.

In the classical theory we found a unique supersymmetric vacuum located at the origin of field space. We will now show that non-perturbative contributions lift the vacuum energy to a positive value, breaking supersymmetry.

As discussed in section \ref{Sec:DFlat}, along the D-flat directions the gauge symmetry is broken as follows
\[
\begin{tikzcd}
	SU(M+4)_1\times SU(M)_2 \times SU(M+4)_3 \arrow[d, "v"] \\
	 SU(5)_1 \times SU(5)_3 \times USp(M-1)_\text{diag}
\end{tikzcd}
\]
where we have assumed that all the VEVs $v_i $ are equal to $v$, as in \eqref{theory_pf_directions_2}. We will justify this assumption at the end of this section, by noting that the vacuum energy density is a convex function of the D-flat moduli implying that its minimum correspond to all VEVs being equal. 

At low energies, the theory reduces to the Twin $SU(5)$ model, which breaks supersymmetry with a vacuum energy given by 
\begin{equation}
	V_{\text{Twin-}SU(5)} \sim \Lambda_{L1}^4 + \Lambda_{L3}^4 \, ,
\label{vacuum_energy_twin_model}
\end{equation}
where the $\sim$ indicates an order of magnitude for each term.

The next step is to connect the dynamical scales $\Lambda_{L1,L3}$ of the low energy Twin $SU(5)$ model to the high energy dynamical scales 
$\Lambda_{1,3}$. In other words, we want to relate the dynamical scales of gauge groups 1 and 3 for energies much lower and much higher than $v$. We do so by scale matching at $E=v$, requiring that
\begin{equation}
	\left(\frac{\Lambda_{L1}}{v}\right)^{\beta_{L1}} =	\left(\frac{\Lambda_{1}}{v}\right)^{\beta_1} ~~,~~\left(\frac{\Lambda_{L3}}{v}\right)^{\beta_{L3}} =	\left(\frac{\Lambda_{3}}{v}\right)^{\beta_3}  \, .
\label{scale_matching}
\end{equation}
Here $\beta_{1,3}$ and $\beta_{L1,L3}$ are the beta function coefficients for gauge groups 1 and 3 at energies higher and lower than $v$, respectively,  
which read
\begin{equation}
\beta_{1} = \beta_{3} = 2M + 11~~,~~ \beta_{L1} = \beta_{L3} = 13~.
\end{equation}
From the relations \eqref{scale_matching}, we obtain the IR dynamical scales in terms of the UV ones and $v$. Plugging these expressions into \eqref{vacuum_energy_twin_model}, we get
\begin{equation}
	V_{\text{Twin-}SU(5)} \sim v^{-4(\beta_1/\beta_{L1} - 1)}\left(\Lambda^{4\beta_1 / \beta_{L1}}_1 + \Lambda^{4\beta_1 / \beta_{L1}}_3\right)\, ,
\end{equation}
where we have used that $\beta_1=\beta_3$ and $\beta_{L1}=\beta_{L3}$. From now on, we use the notation
\begin{equation}
\Lambda^{4\beta / \beta_{L}} \equiv	\left(\Lambda^{4\beta / \beta_{L}}_1 + \Lambda^{4\beta / \beta_{L}}_3\right) \, , \quad \text{where} \quad \beta \equiv \beta_1 \, , \quad \beta_{L} \equiv \beta_{L1} \, .
\end{equation}
Adding the contribution from the perturbative superpotential \eqref{perturbative_superpotential_pseudoflat_directions}, the vacuum energy density becomes
\begin{equation}
	V(v, \Lambda) \sim \vert h \vert^2 v^{10} +  v^{-4(\beta/\beta_{L} - 1)}\Lambda^{4\beta / \beta_{L}} \, . \label{Eq:EnM}
\end{equation}
This function is minimized at a non-zero value for $v$
\begin{equation}
	v_\text{min} \propto \left(\frac{4(\beta/\beta_{L} - 1) }{10}\frac{1 }{\vert h \vert^2 \Lambda^6} \right)^{\frac{1}{6+4\beta/\beta_{L}}} \Lambda  = \left( \frac{8}{130}\frac{M-1 }{\vert h \vert^2 \Lambda^6} \right)^{\frac{13}{8M + 122}} \Lambda  \, .
\end{equation}
The vacuum energy density becomes
\begin{equation}
	\begin{array}{rcl}
		V_\text{min} &\propto& \left(1 + \frac{10}{4(\beta/\beta_L -1)} \right)\left(\frac{4(\beta/\beta_L - 1)}{10} \right)^{\frac{10}{6+4\beta/\beta_L}} \left(\vert h \vert^2 \Lambda^6\right)^{4\frac{\beta/\beta_L -1}{6+4\beta/\beta_L}} \Lambda^4 \, , \vspace{0.1cm}\\
		&=& \left(1 + \frac{130}{8(M -1)} \right)\left(\frac{8(M - 1)}{130} \right)^{\frac{130}{8M+122}} \left(\vert h \vert^2 \Lambda^6\right)^{\frac{8M-8}{8M+122}} \Lambda^4 \, ,
	\end{array}
\end{equation}
which is non-zero, implying that supersymmetry is dynamically broken.\footnote{Notice that this expression consistently gives $V\propto \Lambda^4$ in the $M\to 1$ limit, which corresponds to the SUSY breaking in the Twin $SU(5)$ model \cite{Argurio:2020dkg}.}

\paragraph{Convexity of the vacuum energy density.} 
The calculation above assumed that all VEVs parametrizing D-flat directions were equal, $v_i=v$. We now justify why this is a meaningful assumption. Consider a generic set of VEVs and, without loss of generality, assume that $v_1 \geqslant v_2 \geqslant \cdots \geqslant v_{(M-1)/2}>0$. This hierarchy of scales results in the following higgsing pattern
\[
\begin{tikzcd}
	SU(M+4)_1\times SU(M)_2 \times SU(M+4)_3 \arrow[d, "v_1"] \\
	SU(M+2)_1 \times SU(M-2)_2 \times SU(M+2)_3 \times USp(2)_\text{diag} \arrow[d, "v_2"] \\
	\vdots \arrow[d, "v_{(M-1)/2}"] \\
	SU(5)_1 \times SU(5)_3 \times USp(2)_\text{diag}^{(M-1)/2}
\end{tikzcd}
\]
Performing a sequence of scale matchings, we find the vacuum energy density
\begin{equation}
	\begin{array}{rcl}
	V(v_1, v_2, \cdots, v_{(M-1)/2}, \Lambda) &\sim& \vert h \vert^2 (v_1^{10} +v_2^{10}+ \cdots + v_{(M-1)/2}^{10} ) \vspace{0.2cm}  \\
	& & +  v_1^{-16/\beta_{L}} v_2^{-16/\beta_{L}} \cdots v_{(M-1)/2}^{-16/\beta_{L}} \, \Lambda^{4\beta / \beta_{L}} \, .	
	\end{array}
\end{equation}
This function is symmetric under the permutations of the $v_i$'s and is convex on its domain. Therefore, its minimum must correspond to all the $v_i$'s being equal.

\subsection{Runaway for even $M$}
\label{section_even_M}

So far our analysis assumed $M$ to be odd. In this section we discuss what changes when $M$ is even. We will see that in this case the theory exhibits a runaway behavior instead of supersymmetry breaking in a stable vacuum.

Let us first consider $M=0$. In this case, the gauge theory is a double copy of $SU(4)$ with an antisymmetric tensor $A$. Each gauge group generates an ADS superpotential
\begin{equation}
	W_{\text{np}} = \frac{\Lambda^{11/3}}{\mathrm{Pf} A^{1/3}} \, ,
\end{equation} 
such that the vacuum breaks supersymmetry in a runaway fashion. Note that classically, a VEV of $A$ generically breaks $SU(4)$ to $USp(4)$ (this is the same as saying that the VEV of a fundamental of $SO(6)$ breaks it to $SO(5)$). The superpotential above is then obtained straightforwardly by scale matching and symmetry arguments. 

Reference \cite{Affleck:1984xz} analyzes a closely related model, which is nothing but our model with even $M$, but with $SU(M)_2$ being a global flavor group. The vacuum is again runaway because the superpotential does not lift all classically flat directions. Below, we argue that this conclusion remains true after gauging $SU(M)_2$.

For even $M$, we find four additional gauge invariants with respect to the odd case
\begin{equation}
	\begin{array}{ccl}
		\mathrm{Pf} A_{11} &=& \epsilon_{a_1 \cdots a_{M+4}} (A_{11})^{a_1 a_2} \cdots (A_{11})^{a_{M+3} a_{M+4}} \, , \\[.1 cm]
		\mathcal{Y}_1  &=&  (A_{11})^{a_1 a_2} \cdots (A_{11})^{a_{M-1} a_M}(X_{12})_{a_1}^{\,\,\, i_1}\cdots(X_{12})_{a_{M}}^{\,\,\, i_M} \epsilon_{i_1 \cdots i_{M}} \, , \\[.1 cm]
		\mathcal{Y}_3  &=& \epsilon^{i_1 \cdots i_{M}}(X_{23})_{i_1}^{\,\,\, m_1}\cdots(X_{23})_{i_{M}}^{\,\,\, m_M} (A_{33})_{m_1 m_2} \cdots (A_{33})_{m_{M-1} m_M} \, , \\[.1 cm]
		\mathrm{Pf} A_{33} &=& \epsilon^{m_1 \cdots m_{M+4}} (A_{33})_{m_1 m_2} \cdots (A_{33})_{m_{M+3} m_{M+4}} \, .
	\end{array} 
\end{equation}
Using the relation $\epsilon_{i_1 \cdots i_{M}}\epsilon^{j_1 \cdots j_{M}}=M!\  \delta^{[j_1}_{i_1}\dots \delta^{j_M]}_{i_M}$, one can show that the product $\mathcal{Y}_1 \mathcal{Y}_3$ can be expressed as a sum of products of the gauge invariants $\mathrm{tr}\mathcal{X}^k$. The latter are all set to zero by the F-term conditions when the superpotential is taken into account, leaving us with the condition $\mathcal{Y}_1 \mathcal{Y}_3=0$. So we are left with three flat directions  parametrized by $\mathrm{Pf}A_{11}$, $\mathrm{Pf}A_{33}$ and either $\mathcal{Y}_1$ or $\mathcal{Y}_3$. 

Quantum corrections can be determined by first neglecting the effect of the superpotential. We find that $SU(M+4)_1 \times SU(M)_2 \times SU(M+4)_3 $ is generically broken down to $USp(4)_1 \times USp(4)_3 \times USp(2)_{\text{diag}}^{M/2}$. Gaugino condensates for each of these gauge factors contribute a $\Lambda_L^3$ to the superpotential at low energy. Using scale matching, we find that the term coming from $USp(4)_1$ can be expressed in terms of the dynamical scale $\Lambda_1$ of $SU(M+4)_1$ as follows
\begin{equation}
	W_{\text{np},1} = \left(\frac{\Lambda_1^{2M+11}}{\mathrm{Pf}A_{11} \, \mathcal{Y}_1}\right)^{1/3}\, ,
\end{equation}
and similarly for $USp(4)_3$ and $SU(M+4)_3$. On the other hand, the contributions from the diagonal $USp$ only depend on the product of $\mathcal{Y}_1$ and $\mathcal{Y}_3$.

Combining the tree-level and non-perturbative contributions to the superpotential, we see that the Pfaffians only appear with negative powers. This implies a runaway along the $\mathrm{Pf}A_{11}$ and $\mathrm{Pf}A_{33}$ directions.

\section{Comments on the geometric dual}
\label{geomdual}

In gauge/gravity duality any given gravitational background is dual to a specific vacuum of the corresponding dual gauge theory. For non-conformal theories,  this background has typically a non-trivial radial evolution and accounts for the dynamics along the RG flow and of its endpoint. As such, holography provides a dual geometric description of field theory phenomena, typically in terms of deformations of the dual geometry, brane dynamics or a combination of both.

For example, confinement in pure SYM is related to a complex structure deformation of the dual geometry, the volume of the blown-up 3-cycle being proportional to the gaugino condensate \cite{Klebanov:2000hb,Maldacena:2000yy}. The exact pattern of chiral symmetry breaking in the gauge theory can be understood in terms of the transformation properties of the dual supergravity background under bulk gauge symmetries \cite{Bertolini:2001qa,Klebanov:2002gr,Bertolini:2002xu} (see \cite{Apruzzi:2021phx} for a rephrasing using one-form symmetry arguments). Coulomb branch dynamics is described by ${\cal N}=2$ fractional branes exploring non-isolated singularities  \cite{Bertolini:2000dk,Polchinski:2000mx}, a large-$N$ version of the Seiberg-Witten curve is described via the so-called enhan\c{c}on mechanism \cite{Johnson:1999qt} while ${\cal N}=2$ supergravity backgrounds with varying fluxes find an interpretation as baryonic root transitions in the dual gauge theory \cite{Benini:2008ir} (see also \cite{Polchinski:2000mx,Aharony:2000pp} for previous attempts). 
 
It is then natural to ask what is the geometric dual description of stable dynamical supersymmetric breaking in the Octagon model. A mathematically rigorous answer to this question would require an algebro-geometric analysis of the Octagon singularity which is beyond the scope of the present paper, and is left to future work. However, as we argue below, the field theory analysis of section \ref{DSB_M} and what is known about D-brane dynamics in these contexts, hint for what could be the scenario for the stabilization mechanism in the bulk and the corresponding geometry. 

Dynamical supersymmetry breaking into runaway vacua is due to a geometric obstruction against complex structure deformations \cite{Berenstein:2005xa,Franco:2005zu}. In supersymmetric confining theories, like the conifold duality cascade \cite{Klebanov:2000hb}, the magnetic RR-flux associated to the fractional branes is supported by a 3-cycle that blows-up so that its volume stabilizes to a value determined by the flux, {\it i.e.} by the number of fractional branes. When the complex structure deformation is obstructed, this stabilization cannot occur and the singular geometry becomes effectively repulsive. More technically, the F-term equations of certain baryonic operators set the parameter of the complex structure deformation to be proportional to $\Lambda^\#/X$, with $X$ a gauge invariant operator. However, since the complex structure deformation is obstructed this implies that the parameter has to vanish, and the F-term equation can be satisfied only at $X \rightarrow \infty$ (see \cite{Argurio:2007vq}  for a detailed discussion for the prototype CY where this happens, namely the complex cone over the first del Pezzo surface). 

As far as fluxes are concerned, orientifolds behave very much like fractional branes. 
So, generically, orientifolds break conformal invariance and induce an RG flow in the dual field theory, the fate of the RG flow depending on the specific CY singularity and brane content under consideration \cite{Argurio:2017upa,Antinucci:2020yki,Antinucci:2021edv,Amariti:2021lhk}. In the case of the Octagon, while $N$ regular branes probing it describe a SCFT, in the presence of an orientifold the theory breaks supersymmetry and the vacuum becomes runaway \cite{Argurio:2020dkg}. So in this case the orientifold plane behaves as a DSB fractional brane. 

As we have seen, adding to this configuration an odd number $M$ of deformation fractional branes  the runaway is cured: supersymmetry is broken but the vacuum is stabilized at finite distance in field space. What is happening, geometrically?

The geometric effect of this class of fractional branes is to trigger a complex structure deformation on the CY, blowing-up the 3-cycle dual to the 2-cycle they wrap. From figure \ref{Fig:Octagon} one can see that our fractional branes share some (but not all) faces in the dimer with the orientifold.\footnote{More precisely, the faces of the dimer are populated by the (anomalous) fractional branes that need to be added in order to cancel the orientifold tadpole in the directions in which the dual cycle is compact. Here and in the following by `orientifold' we denote all of these objects together.} This means that the orientifold and the fractional branes are wrapping cycles inside the CY that partially overlap and so do their dual 3-form fluxes. Following previous discussion, this suggests that the geometric transition could allow the orientifold flux to stabilize and, in turn, smooth out the geometry. 

The above observations do not seem to distinguish between $M$ being even or odd, while the field theory analysis shows a sharp distinction between the two cases.  So one might wonder how such difference emerges, geometrically. Let us start considering $M$ odd.

The Twin $SU(M+4)$ model describing the gauge theory dynamics at the bottom of the cascade is higgsed, in this case, to a Twin $SU(5)$ model together with a decoupled pure $USp(M-1)$ gauge factor, that is
\begin{eqnarray}
\label{sbp1}
	SU(M+4)_1\; \times &SU(M)_2& \times \;SU(M+4)_3 \nonumber \\ 
	&\downarrow v & \\ 
	 SU(5)_1\; \times &SU(5)_3& \times \; USp(M-1) \nonumber
\end{eqnarray}
This suggests that what happens geometrically is that the initial stack of $M$ fractional branes gets separated into two,  at a relative distance $v \sim \Lambda$. This VEV is a quantum effect, and this agrees with the fact that within the undeformed Octagon geometry, deformation fractional branes cannot move away from the tip. The $USp$ factor in \eqref{sbp1} suggests that the brane separation happens along the orientifold plane, and so the stack must necessarily be made of an even number of them. This is why, for $M$ odd, one gets a (Twin) $SU(5)$ model. As already emphasized, fractional branes induce a complex structure deformation, the volume of the blown-up 3-cycle being proportional to the strong coupling scale $\Lambda$ of the Twin $SU(M+4)$ model and, by scale matching, to those of the confining $USp$ factor and the supersymmetry breaking Twin $SU(5)$ model (and hence to the non-vanishing vacuum energy). Note that as the deformation branes are self-identified under the orientifold action, it is expected that the same is true for the 3-cycle that blows up. 

As discussed in section \ref{DSB_M}, for even $M$ the higgsing pattern is different and the supersymmetry breaking dynamics is that of a Twin $SU(4)$ model, which is runaway. In this case the vacuum energy is minimized at infinity in field space 
and the geometry is expected to be singular \cite{Franco:2005zu}. How can this happen, geometrically? As shown in section \ref{section_even_M}, the runaway in the even $M$ case is associated to the presence of some baryonic (Pfaffian) operators which are absent in the $M$ odd case. As reviewed in section \ref{review Oct}, baryonic operators are related to (possibly non-toric) partial resolutions which open up directions along which the branes can now slide away. 
It is then tempting to speculate that the single fractional brane stuck at the origin provides an obstruction to such resolution, which is instead allowed when $M$ is even. This is not surprising because it is known that adding a brane on top of an orientifold changes the nature of the latter \cite{Witten:1998xy,Hanany:2000fq}.

As a final comment, let us recall that stability, when it occurs, is due to a balance between two effects which are of the same order. This is something we also see in our set-up. The orientifold charge is order one in units of fractional brane charge, and so is the charge of the one fractional brane that balances it. At the same time the curvature of the background is small in string units, since the volume of the 3-cycle is proportional to $M$, which is large. 

\newpage
\section{Discussion}
\label{disc}

In this paper we have shown that the dynamical supersymmetry breaking model of \cite{Argurio:2020dkg,Argurio:2020npm} admits a large $M$ generalization. By the very meaning of the gauge/gravity duality, this implies that there exists a dual ten-dimensional type IIB supergravity background which breaks supersymmetry and is stable. This opens up the possibility of having a weakly coupled gravity dual which can capture the dynamics of a cascading supersymmetric gauge theory from the UV all the way down to the deep IR and which admits stable supersymmetry breaking vacua at its bottom. 

Our result fills a long-searched missing entry in the AdS/CFT duality dictionary, showing that also stable non-supersymmetric dynamics can be described using a geometric dual and that this is possible within a top-down approach. We believe that more interesting lessons will be learned by a deeper understanding of the geometric structure of the Octagon as well as by generalizing to other toric CY singularities the same results, following the strategy outlined in \cite{Argurio:2020npm}. 

We would like to comment on the implications of our results in the wider context of the string landscape and the Swampland program. While we are working in the strict gauge/gravity regime and so the effective four-dimensional Planck mass vanishes, one might wonder whether the existence of a non-supersymmetric ten-dimensional gravity background could provide a counterexample to any of the recently proposed Swampland conjectures. 

The structure of the ten-dimensional background would be similar to the KS solution \cite{Klebanov:2000hb}  and more generally to any background with varying fluxes: 
a warped throat which has log-corrections with respect to an asymptotically AdS spacetime, and with a 5-form flux increasing towards the boundary. The difference here is that supersymmetry is broken at the bottom of the throat. As such, the Octagon model can be seen as a counterexample to the conjecture forbidding stable non-supersymmetric locally AdS warped throats proposed in \cite{Buratti:2018onj}. On the other hand, it does not contradict the (sharpened version) of  the Weak Gravity Conjecture proposed in \cite{Ooguri:2016pdq,Freivogel:2016qwc}, which implies that any non-supersymmetric anti-de Sitter vacuum supported by fluxes must be unstable. Indeed, at least one of the assumptions in \cite{Ooguri:2016pdq,Freivogel:2016qwc}, {\it i.e.} that the solution is asymptotically AdS (and hence dual to a proper CFT), is not met here. As extensively discussed in the literature, see {\it e.g.} \cite{Aharony:2005zr,Buchel:2005cv,Aharony:2006ce}, cascading backgrounds present a few sharp differences with respect to asymptotic AdS ones. In particular there is no well-defined UV fixed point and holographic renormalization should be done with care \cite{Bertolini:2015hua,Krishnan:2018udc}. This is an aspect worth investigating further, given the relevance it could have in sharpening our understanding of the Swampland and, so, of the string landscape.
 
It has further been argued that the absence of global symmetries in quantum gravity implies that any two vacua are connected by a domain wall \cite{McNamara:2019rup}. It is a consequence of this conjecture that any non-supersymmetric compactification can, potentially, decay to a supersymmetric one by nucleation of suitable domain walls. While such domain walls are sometimes conventional objects such as D-branes or bubbles of nothing \cite{GarciaEtxebarria:2020xsr}, they may also be non-supersymmetric objects yet to be discovered \cite{McNamara:2019rup}. We cannot exclude the existence of such an exotic instability that may not be amenable to study with usual QFT tools. For a discussion on these domain walls in the context of AdS/CFT, see \cite{Ooguri:2020sua}.

Supersymmetry breaking D-brane configurations also play a prominent role within de Sitter constructions in string theory. In both most popular scenario, KKLT \cite{Kachru:2003aw} and LVS \cite{Balasubramanian:2005zx},  anti-D-branes placed at the tip of warped throats are a key ingredient \cite{Kachru:2002gs}. The Octagon could furnish an alternative to such ingredient, in the spirit of \cite{Retolaza:2015nvh}. One difference is that the theory breaks supersymmetry in a stable vacuum rather than a metastable one, as in \cite{Kachru:2002gs}. Another difference is that the D-brane configuration giving rise to DSB is supersymmetric and no addition of explicit breaking sources is needed. Both these aspects suggest that using this type of model might lead to somewhat more control. Finally, a construction based on a model where the background already breaks supersymmetry without the need of adding explicit supersymmetry breaking sources to uplift the vacuum energy after moduli are stabilized, could suggest a one-step construction with respect to the two steps characterizing both KKLT and LVS scenario. One might wonder if this could challenge recent criticism on the two-step KKLT construction \cite{Lust:2022lfc}. Other aspects characterizing de Sitter constructions within the aforementioned scenario are not different, like issues related to obtaining a sufficiently small cosmological constant in the effective four-dimensional effective theory after full moduli fixing. We believe that this is yet another aspect worth investigating further, giving the relevance this could have in this context.
	
\section*{Acknowledgements}

We thank Simone Blasi, Shamit Kachru, Miguel Montero, and Thomas van Riet for useful discussions. 
R.A. and E.G.-V. acknowledge support by IISN-Belgium (convention 4.4503.15) and by the F.R.S.-FNRS under the ``Excellence of Science'' EOS be.h project n.~30820817. R.A. is a Research Director of the F.R.S.-FNRS (Belgium). M.B. is partially supported by the MIUR PRIN Grant 2020KR4KN2 ``String Theory as a bridge between Gauge Theories and Quantum Gravity'' and by INFN Iniziativa Specifica ST\&FI. The research of S.F. was supported by the U.S. National Science Foundation grants PHY-1820721, PHY- 2112729 and DMS-1854179. The work of S.M. is funded by the scholarship granted by ``Fondazione Angelo Della Riccia.'' The research of A.P. is supported by a Fellowship of the Belgian American Educational Foundation.

\appendix

\section{Space of D-flat directions - details}
\label{section_low_E_theory_generic_vevs}

This appendix provides further details to the discussion in section \ref{Sec:DFlat}. For completeness, let us first rewrite the D-term equations
	\begin{align}
SU(M+4)_1  :  \ \ \ & 2(A_{11})^{bc}(A_{11}^\dagger)_{ca} -  (X_{12}^\dagger)^{\, b}_{i}(X_{12})^{\, i}_{ a} =   \alpha_1  \, \delta^b_{\, a} \, , \label{EqA:Dterm1} \\
SU(M)_2	 :  \ \ \ &  (X_{12})^{\,\, j}_{ a}(X_{12}^\dagger)^{\, a}_{i} - (X_{23}^\dagger)^{\,\, j}_{m}(X_{23})^{\, m}_{ i}  =  \alpha_2  \, \delta^j_{\, i} \, , \label{EqA:Dterm2} \\
SU(M+4)_3 : \ \ \ & (X_{23})^{\, n}_{i}(X_{23}^\dagger)^{\,\, i}_{ m} - 2(A_{33}^\dagger)^{np}(A_{33})_{pm}  =  \alpha_3  \, \delta^n_{\, m} \, . \label{EqA:Dterm3}
	\end{align}
We now exploit the (global part of the) gauge and global symmetries, reported in \Cref{Tab:ClassicalCharges}, to simplify the solutions to these equations. 
\begin{table}
		\centering
		\begin{tabular}{c|ccc|cccc}
			\toprule
			& $SU(M+4)_1$  & $SU(M)_2$ & $SU(M+4)_3$ & $U(1)_1$ &  $U(1)_2$ & $U(1)_3$ & $U(1)_R$ \\
			\midrule
			$A_{11}$ &  $\asymm$ & $\mathbf{1}$ & $\mathbf{1}$ & $2$ & $0$ & $0$ & $1/3$ \\
			$X_{12}$ &$\antifund$ &$\fund$ & $\mathbf{1}$ & $-1$ & $1$ & $0$ & $1/3$ \\
			$X_{23}$ & $\mathbf{1}$ &$\antifund$ &$\fund$ & $0$ & $-1$ & $1$ & $1/3$ \\
			$A_{33}$ &  $\mathbf{1}$ &   $\mathbf{1}$ &$\antiasymm$ & $0$ & $0$ & $-2$ & $1/3$ \\
			\bottomrule
		\end{tabular}
		\caption{Classical symmetries and charges of Twin $SU(M+4)$ model.}
		\label{Tab:ClassicalCharges}
\end{table}
	
Since $SU(M)_2$ only has (anti)fundamental matter, its D-term equations \eqref{EqA:Dterm2} are those of SQCD with $N_f=M+4$ and its analysis is rather standard. Using $SU(M+4)_1\times SU(M)_2$ transformations, we can take $X_{12}$, {\it i.e.} the `quarks' of gauge group 2, to block diagonal form
	\begin{equation}
	X_{12}  = \left(\begin{array}{cccc}
			w_1 & 0 & \cdots & 0 \\
			0 & w_2 & \cdots & 0\\
			\vdots & \vdots  & \ddots & \vdots \\
			0 & 0 &  \cdots & w_M \\
			0 & 0 &  \cdots & 0  \\
			0 & 0 &  \cdots & 0 \\
			0 & 0 &  \cdots & 0 \\
			0 & 0 &  \cdots & 0  \\
		\end{array}\right) \, . \label{Eq:X12}
	\end{equation}
For now, we allow the non-zero diagonal entries to take complex values.  
This leaves us with a $U(1)^{M}_1 \times SU(4)_1\times U(1)^{M-1}_2$ gauge symmetry to be used later. Let us now focus on $X_{23}$, {\it i.e.} on the `antiquarks' of gauge group 2. Using $SU(M+4)_3$, we can simultaneously lower-diagonalize it to the form
\begin{equation}
	X_{23}  = \left(\begin{array}{cccccccc}
		\tilde{w}_1 & 0 & \cdots & \ 0 \ \ & \ 0 \ \ & \ 0 \ \ & \ 0 \ \ & \ 0 \ \ \\
		\tilde{w}_{21} & \tilde{w}_2 & \cdots & 0& 0& 0& 0& 0\\
		\vdots & \vdots  & \ddots & \vdots & \vdots & \vdots & \vdots & \vdots \\
		\tilde{w}_{M1} & \tilde{w}_{1M} &  \cdots & \tilde{w}_M & 0& 0& 0& 0\\
	\end{array}\right) \, .
	\label{Eq:X23}
\end{equation}
For complex values on the diagonal, the leftover gauge symmetry is $U(1)^{M}_3 \times SU(4)_3$. Using \eqref{Eq:X12} and \eqref{Eq:X23}, \eqref{EqA:Dterm2} becomes
	\begin{equation}
	\begin{array}{rcll}
	\vert w_i \vert^2 - \vert \tilde{w}_i \vert^2 &=& \alpha_2 & \quad \text{for every }i\, , \\
	\tilde{w}_{ij} &=&  0 & \quad \text{ for }i \neq j \, .
	\end{array}
	\end{equation}
	
Using \eqref{Eq:X12}, equation \eqref{EqA:Dterm1} tells us that $A_{11} A_{11}^\dagger$ is diagonal. Moreover, the antisymmetry of $A_{11}$ implies that $A_{11} A_{11}^\dagger$ has pairs of equal and positive eigenvalues. Since $X^\dagger_{12} X_{12}$ is of rank $M$ odd and also diagonal with positive eigenvalues, a bit of algebra shows that $A_{11} A_{11}^\dagger$ has at most $M-1$ non-negative entries and $\alpha_1=0$. An example of a solution, up to $\mathbb{Z}_{M}$ permutations, is
	\begin{equation}
		A_{11}A_{11}^\dagger = \frac{1}{2} \left(\begin{array}{cccc|c}
			\vert w_1 \vert^2 \mathbb{1}_{2\times2} & \mathbb{0}_{2\times2} & \cdots & \mathbb{0}_{2\times2} & \mathbb{0}_{2\times5} \\
			\mathbb{0}_{2\times2} &  \vert w_3\vert^2 \mathbb{1}_{2\times2} & \cdots & \mathbb{0}_{2\times2} & \mathbb{0}_{2\times5} \\
			\vdots & \vdots  & \ddots & \vdots  & \vdots  \\
			\mathbb{0}_{2\times2} & \mathbb{0}_{2\times2} & \cdots &  \vert w_{M-2}\vert^2 \mathbb{1}_{2\times2} & \mathbb{0}_{2\times5} \\ \hline
			\mathbb{0}_{5\times2} & \mathbb{0}_{5\times2} & \cdots &  \mathbb{0}_{5\times2} & \mathbb{0}_{5\times5} \\
		\end{array} \right)\, , \label{Eq:AAdag}
	\end{equation}
	implying that
	\begin{equation}
		\vert w_1 \vert = \vert w_2 \vert  \, , \quad \cdots \, , \quad \vert w_{M-2} \vert = \vert w_{M-1} \vert \, , \quad \vert w_M \vert = 0 \, .
	\end{equation}
We can use $U(1)^{(M-1)/2}_1$ to align the phase of each $w_{2i}$ with the one of $w_{2i-1}$. Then, we define
\begin{equation}
	v_i \equiv w_{2i-1} \, ,
\end{equation}
for $i=1, \cdots, (M-1)/2$. This brings $X_{12}$ to the form
\begin{equation}
X_{12} = \left(\begin{array}{cccc|c}
	v_1 \mathbb{1}_{2\times2}  & \mathbb{0}_{2\times2} & \cdots & \mathbb{0}_{2\times2} & \mathbb{0}_{2\times1} \\
	\mathbb{0}_{2\times2} & v_2 \mathbb{1}_{2\times2}  & \cdots & \mathbb{0}_{2\times2} & \mathbb{0}_{2\times1} \\
	\vdots & \vdots  & \ddots & \vdots & \vdots \\
	\mathbb{0}_{2\times2}  & \mathbb{0}_{2\times2}  &  \cdots & v_{(M-1)/2} \mathbb{1}_{2\times2} &  \mathbb{0}_{2\times1}  \\ \hline
	\mathbb{0}_{5\times2} & \mathbb{0}_{5\times2} &  \cdots & \mathbb{0}_{5\times2} & \mathbb{0}_{5\times1}   \\
\end{array}\right) \, .\label{Eq:X12n}
\end{equation}	
The only solution\footnote{\color{black}Indeed, the expression for $A_{11}A_{11}^\dagger$ in \eqref{Eq:AAdag} is solved by any $(A_{11})_{ab}= (\mathbf{e}_a)_b$, where $\lbrace\mathbf{e}_a\rbrace$ is an orthogonal basis of vectors in $\mathbb{C}^{M-1}$, where each pair $\lbrace \mathbf{e}_{2a-1}, \mathbf{e}_{2a} \rbrace$ shares the same norm, 
 and at the condition of showing antisymmetry under $a\leftrightarrow b$. Starting from the solution for $A_{11}$ shown in \eqref{Eq:A11}, we can reach any other orthogonal basis by performing a unitary transformation $A_{11}\rightarrow U A_{11} U^{\dagger}$, but the latter is antisymmetric only for $U^\dagger =U^{T}$, and the norms are preserved only if $U\in U(2)\times \cdots \times U(2)$. This means that $A_{11}$ can be transformed under a subset of $SU(2)\times \cdots \times SU(2)$, but $A_{11}$ is actually preserved by this group, so we conclude that eq.~\eqref{Eq:AAdag} is the unique solution.} to \eqref{Eq:AAdag} is 
\begin{equation}
 A_{11} = \frac{1}{\sqrt{2}} \left(\begin{array}{cccc|c}
 	v_1 \mathbb{J}_{2\times2} & \mathbb{0}_{2\times2} & \cdots & \mathbb{0}_{2\times2} & \mathbb{0}_{2\times5} \\
 	\mathbb{0}_{2\times2} &   v_2 \mathbb{J}_{2\times2} & \cdots & \mathbb{0}_{2\times2} & \mathbb{0}_{2\times5} \\
 	\vdots & \vdots  & \ddots & \vdots  & \vdots  \\
 	\mathbb{0}_{2\times2} & \mathbb{0}_{2\times2} & \cdots &   v_{(M-1)/2} \mathbb{J}_{2\times2} & \mathbb{0}_{2\times5} \\ \hline
 	\mathbb{0}_{5\times2} & \mathbb{0}_{5\times2} & \cdots &  \mathbb{0}_{5\times2} & \mathbb{0}_{5\times5} \\
 \end{array} \right)\, , \label{Eq:A11}
 \end{equation}
where we used another $U(1)^{(M-1)/2}_1$ to align the phases with those of $X_{12}$, and $(\mathbb{J}_{2 \times 2})^{ab}=\epsilon^{ab}$.

The vanishing of $w_M$ back-reacts onto \eqref{EqA:Dterm2} and implies that $\alpha_2 =0$. Hence,
	\begin{equation}
		\vert v_i \vert = \vert \tilde{w}_{2i-1}\vert \, .
	\end{equation}
We can use the gauge $U(1)^{M-1}_2$ and the global $U(1)_2$ transformations to set 
\begin{equation}
X_{23} = X_{12}^T \, .
\label{X23=X12T}
\end{equation}	
Finally, \eqref{EqA:Dterm3} leaves no other choice than imposing $\alpha_3 = 0$, and we find
\begin{equation}
		A_{33}^\dagger A_{33} = A_{11}A_{11}^\dagger \, .
\end{equation}
The solution, using a $U(1)^{(M-1)/2}_3$ rotation, becomes
\begin{equation}
	A_{33} = A_{11} \, .
	\label{A33=A11}
\end{equation}

Now that the phases are all aligned and are the same in each matrix for the UV fields, we can perform an ultimate $U(1)^{(M-1)/2}_3$ transformation to set all the $v_i$'s to real positive values. The unused mix of global and gauge transformations is $U(1)^2_1\times SU(4)_1 \times U(1)^2_3\times SU(4)_3 $.\footnote{The alert reader might notice that this symmetry is similar, but not exactly equal to the one after the higgsing associated to the VEVs at a point in the moduli space, to be discussed later (see e.g. \eqref{theory_pf_directions_1} for the case of different $v_i$'s). This is because the vanishing of the D-terms forces some combinations of the VEVs to vanish, resulting in an enhanced symmetry.}  

In summary, combining \eqref{Eq:X12n}, \eqref{Eq:A11}, \eqref{X23=X12T} and \eqref{A33=A11}, we have shown that, in the absence of superpotential, the moduli space is parametrized by the $(M-1)/2$ parameters $v_i$. 

Let us now study the low energy effective dynamics on the space of D-flat directions. 

\paragraph{Generic vacuum expectation values.} 
For generic VEVs of the form \eqref{Eq:X12n} and \eqref{Eq:A11}, the gauge symmetry is 
	\begin{equation}
		SU(5)_1 \times SU(1)_2 \times SU(5)_3 \times SU(2)_\text{diag}^{\frac{M-1}{2}} \, ,
		\label{gauge_symmetry_generic_vevs}
	\end{equation}
	where $SU(2)_\text{diag}^{\frac{M-1}{2}}$ indicates the product
	\begin{equation}
		\underbrace{SU(2) \times \cdots \times SU(2)}_{\frac{M-1}{2} \text{ times}}
	\end{equation}
	diagonally embedded in
	\begin{equation}
		SU(M-1)_1 \times SU(M-1)_2 \times SU(M-1)_3 \, .
	\end{equation}
Each of these $SU(2)$ factors acts on one of the $2\times 2$ blocks on the diagonals of the fields in \eqref{Eq:X12n} and \eqref{Eq:A11}.
	
In this generic vacuum, the number of broken gauge generators is
	\begin{equation}
2 \left((M+4)^2 - 1\right) +  \left(M^2 - 1\right)  - 2 \left(5^2 -1\right) - 3\frac{M-1}{2}= 3M^2 + \frac{29}{2} M - \frac{35}{2}\, ,
\end{equation}
and therefore the same number of chiral superfields are eaten. Subtracting this from the initial number of chiral fields we find
\begin{equation}
2 \frac{(M+4)(M+3)}{2} + 2 M(M+4) - 3M^2 - \frac{29}{2} M + \frac{35}{2} = \frac{M-1}{2} + 30\,  \label{Eq:Counting}
\end{equation}
leftover chiral fields, which we identify below.

Consider $X_{12}$ first. It can be written as
\begin{equation}
X_{12} = \left< X_{12} \right> + \delta \tilde{X}_{12} \, , \label{Eq:Pert}
\end{equation}
where $\left< X_{12} \right>$ is given by \eqref{Eq:X12n} and $\delta \tilde{X}_{12}$ is a small perturbation. We need to be more precise and distinguish the physical massless\footnote{We do not include in eq.~\eqref{Eq:Pert} the chiral fields that acquired a mass as a consequence of the super-Higgs mechanism.} degrees of freedom $\delta X_{12}$ from those that can be reached by a gauge transformation around $\left< X_{12} \right>$, {\it i.e.} the eaten Goldstone bosons,
\begin{equation}
\delta\tilde{X} _{12} = \delta {X}_{12} + i \delta\xi^A \lambda^A \left<X_{12}\right> - i \delta\chi^I  \left<X_{12}\right> \mu^I \, ,
\end{equation}
where $\lambda^A$ and $\mu^I$ are generators of $\mathfrak{su}(M+4)$ and $\mathfrak{su}(M)$ respectively, while $\delta \xi^A$ and $\delta \chi^I$ are 
real fields. The criterion for distinguishing $\delta X_{12}$ from $\delta \xi^A$ and $\delta \chi^I$ is that they should be orthogonal to each other, {\it i.e.} fill different entries in the $(M+4)\times M$ matrix.  
Similarly, for $A_{11}$ we have
\begin{equation}
A_{11} = \left< A_{11}\right> + \delta {A}_{11}  + i \delta\xi^A \lambda^A \left<A_{11}\right> + i \delta\xi^A  \left<A_{11}\right> \lambda^{T\, A} \, ,
\end{equation}
where $\left<A_{11}\right>$ is given by \eqref{Eq:A11}. 

By definition, the gauge contributions to the last two equations vanish for generators that preserve the vacuum, which in turn serves as guidance for where to look for physical degrees of freedom. For instance, invariance of the vacuum under the $SU(5)_1$ generators implies that the last $5\times 5$ block in the diagonal of $A_{11}$ cannot be accessed with broken generators and must be a physical $5\times 5$ antisymmetric field, which we denote $\delta a_{11}$. In other words,
\begin{equation}
	\delta	{A}_{11} = \frac{1}{\sqrt{2}}\left(\begin{array}{ccccc|c}
		  & &  &   &  & \\
	 	  & &  &   &  & \\
		  & & (\boldsymbol{?})_{M-1 \times M-1}  &   &  & (\boldsymbol{?})_{M-1 \times 5} \\
		  &  &  &   &  & \\
		  &  &  &   &  & \\ \hline
		  &  & (\boldsymbol{?})_{5 \times M-1} &  & & (\delta a_{11})_{5\times5} \\
	\end{array} \right)\, .
\end{equation}
Similarly, we find physical fields $\delta x_1$, $\delta x_3$, and $\delta a_{33}$ in the bottom-right blocks of $X_{12}$, $X_{23}$, and $A_{33}$ respectively. These fields correspond to the 30 in \eqref{Eq:Counting} and can be organized into representations of $SU(5)_1 \times SU(5)_3$ as follows\begin{equation}
	30 = 10 +5 +5 + 10 \, , 
\end{equation}
corresponding to $\asymm_1$, $\antifund_1$, $\fund_3$, and $\antiasymm_3$ respectively.

Let us now focus on the remaining $(M-1)/2$ chiral fields, which we call $\delta \sigma_{i}$. There is one such field for every $SU(2)_\text{diag}$. 
Hence, it is reasonable to assume that each $\delta \sigma_{i}$ is at most charged under one $SU(2)_\text{diag}$. As supporting evidence, note that all entries of the matrices outside the diagonal blocks can be accessed via a gauge transformation of broken generators (in the case of $A_{11}$, they belong to the coset $SU(M+4)_1 \backslash SU(2)_1\times \cdots \times SU(2)_1\times SU(5)_1$). Hence, the $\delta \sigma_{i}$ must populate the $2\times 2$ blocks. Moreover, due to their dimension, they can only be singlets. Therefore, they do not participate in the gauge dynamics at low energy. 
There exists a gauge fixing in which 
\begin{equation}
	\delta	{A}_{11} = \frac{1}{\sqrt{2}}\left(\begin{array}{cccc|c}
		\delta \sigma_1 \mathbb{J}_{2\times2} & \mathbb{0}_{2\times2} & \cdots & \mathbb{0}_{2\times2} & \mathbb{0}_{2\times5} \\
		\mathbb{0}_{2\times2} &  \delta \sigma_2 \mathbb{J}_{2\times2} & \cdots & \mathbb{0}_{2\times2} & \mathbb{0}_{2\times5} \\
		\vdots & \vdots  & \ddots & \vdots  & \vdots  \\
		\mathbb{0}_{2\times2} & \mathbb{0}_{2\times2} & \cdots &   \delta \sigma_{(M-1)/2} \mathbb{J}_{2\times2} & \mathbb{0}_{2\times5} \\ \hline
		\mathbb{0}_{5\times2} & \mathbb{0}_{5\times2} & \cdots &  \mathbb{0}_{5\times2} & (\delta a_{11})_{5\times5} \\
	\end{array} \right)\, ,
\end{equation}
\begin{equation}
	\delta {X}_{12} = \left(\begin{array}{cccc|c}
		\delta \sigma_1 \mathbb{1}_{2\times2}  & \mathbb{0}_{2\times2} & \cdots & \mathbb{0}_{2\times2} & \mathbb{0}_{2\times1} \\
		\mathbb{0}_{2\times2} & \delta \sigma_2 \mathbb{1}_{2\times2}  & \cdots & \mathbb{0}_{2\times2} & \mathbb{0}_{2\times1} \\
		\vdots & \vdots  & \ddots & \vdots & \vdots \\
		\mathbb{0}_{2\times2}  & \mathbb{0}_{2\times2}  &  \cdots &  \delta \sigma_{(M-1)/2} \mathbb{1}_{2\times2} &  \mathbb{0}_{2\times1}  \\ \hline
		\mathbb{0}_{5\times2} & \mathbb{0}_{5\times2} &  \cdots & \mathbb{0}_{5\times2} & (\delta x_1 )_{5\times1}   \\
	\end{array}\right)  \, , \label{Eq:Intermediate}
\end{equation}
and similar expressions can be written for $\delta X_{23}$ and $\delta A_{33}$.

The low energy theory at a generic point in the D-flat directions is thus composed of $(M-1)/2$ copies of pure $SU(2)$ SYM, $(M-1)/2$ singlets $\delta \sigma_{i}$, and a Twin $SU(5)$ model, as summarized in \eqref{theory_pf_directions_1}.

\paragraph{Identical vacuum expectation values.} 

Let us now investigate the low energy theory in the special case in which all VEVs $v_i$ are equal, namely 
\begin{equation}
v_1 = \cdots = v_{(M-1)/2} \equiv v \, . \label{Eq:PermSym}
\end{equation}

Compared to \eqref{gauge_symmetry_generic_vevs}, the diagonal subgroup is enhanced. While \eqref{Eq:X12n} subject to \eqref{Eq:PermSym} is preserved by an $SU(M-1)_\text{diag}$, \eqref{Eq:A11} imposes a more stringent constraint. We can use an $U \in U(M+4)_1$ transformation to swap rows and columns and write
\begin{equation}
A_{11} \rightarrow U A_{11} U^T = \frac{1}{\sqrt{2}} \left(\begin{array}{c|c}
	v \mathbb{J}_{M-1\times M-1} & \mathbb{0}_{M-1 \times5} \\ \hline
	\mathbb{0}_{5\times M-1} & \mathbb{0}_{5\times5} \\
\end{array} \right)\, .
\end{equation}
It is now clear that the upper-left part of the matrix is preserved by $Sp(M-1,\mathbb{C})$, so the full vacuum structure preserves $USp(M-1)_\text{diag} = Sp(M-1,\mathbb{C}) \cap SU(M-1)$.\footnote{For generic VEVs, we could have written the $(M-1)/2$ $SU(2)_\text{diag}$ as a product of $USp(2)_\text{diag}$, in order to make the group structure more similar to the one that arises for equal VEVs.}

$USp(2)_\text{diag}$ has $M(M-1)/2$ generators, implying that after higgsing we end up with
\begin{equation}
30 + \frac{(M-1)(M-2)}{2}  \label{Eq:CountingSym}
\end{equation}
chiral superfields. The second term in the sum is the dimension of an antisymmetric $(M-1) \times (M-1)$ matrix, indicating that the low energy theory in this vacuum is the Twin $SU(5)$ model plus $USp(M-1)_\text{diag}$ with an antisymmetric tensor $\delta a_\text{d}$. It is convenient to decompose this tensor into a direct sum of its trace
\begin{equation}
	\delta \sigma \equiv \frac{1}{M-1}\mathrm{tr} \mathbb{J}_{M-1 \times M-1} \delta a_\text{d} \, ,
\end{equation}
and a traceless part, that we denote $\delta A_\text{d}$.

The fields $\delta \sigma$ and $\delta A_\text{d}$ appear democratically in all UV matrices, such as
\begin{equation}
{A}_{11} = \frac{1}{\sqrt{2}}\left(\begin{array}{c|c}
		 (v+\delta \sigma) \mathbb{J}_{M-1\times M-1} + \delta A_\text{d}  & \mathbb{0}_{M-1 \times5} \\ \hline
		\mathbb{0}_{5\times M- 1}  & (\delta a_{11})_{5\times5} \\
	\end{array} \right)\, , 
\end{equation}
\begin{equation}
	{X}_{12} = \left(\begin{array}{c|c}
		(v+\delta \sigma) \mathbb{1}_{M-1\times M-1}  + \mathbb{J}_{M-1\times M-1} \delta A_\text{d}^T   & \mathbb{0}_{M-1 \times 1} \\ \hline
		\mathbb{0}_{5\times M- 1}  & (\delta x_{1})_{5\times1} \\
	\end{array} \right)\, , 
\end{equation}
and similar expressions for ${X}_{23}$ and ${A}_{33}$. In this way, we indeed find that, for $U\in USp(M-1)$,
\begin{equation}
X_{12}\rightarrow U^\dagger  \, X_{12}\,  U\quad \text{implies} \quad \delta A_\text{d} \rightarrow U\,  \delta A_\text{d} \, U^T \, .
\end{equation}

In this case, the low energy theory is thus composed of an $USp(M-1)$ SYM, one only singlet $\delta \sigma$, and a Twin $SU(5)$ model, as summarized in \eqref{theory_pf_directions_2}. Note that in the main text we often combine the VEV of $v$ and the dynamical field $\delta \sigma$ into a single dynamical field still denoted $v$, in a slight abuse of notation.


\bibliography{mybib}
\bibliographystyle{JHEP}

\end{document}